\documentclass[12pt]{article}
\usepackage{epsf}
\usepackage{cite}
\newcommand{\sect}[1]{\section{#1}\setcounter{equation}{0}}

\begin{document}
\bigskip
\hspace*{\fill}
\vbox{\baselineskip12pt \hbox{UCSBTH-98-1}\hbox{hep-th/9803085}}
\bigskip\bigskip\bigskip

\centerline{\Large \bf Possible Resolution of Black Hole Singularities}
\bigskip
\centerline{\Large \bf from Large $N$ Gauge Theory}
\bigskip\bigskip\bigskip

\centerline{\large Gary T. Horowitz and Simon F. Ross\footnote{\tt gary,
sross@cosmic.physics.ucsb.edu}}
\medskip
\centerline{Department of Physics,}
\centerline{University of California}
\centerline{Santa Barbara, CA 93106-9530}
\vskip 2cm

\begin{abstract}
  We point out that the recent conjecture relating large $N$ gauge
  theories to string theory in anti-de Sitter spaces offers a
  resolution in principle of many problems in black hole physics.
  This is because the gauge theory also describes  spacetimes
  which are not anti-de Sitter, and include black hole horizons and
  curvature singularities. 
\end{abstract}

\newpage

\sect{Introduction}

A longstanding goal of quantum gravity is to smooth out, or resolve,
the singularities of classical general relativity.  It is now well
known that many spacetimes which are singular from the standpoint of
general relativity are in fact nonsingular when viewed as solutions
to string theory.  An early example of this was orbifold
singularities, in which the curvature remains bounded near the
singular point. This has since been extended to more serious curvature
singularities such as the conifold singularity \cite{strom:coni}.
However, the singularities inside a black hole have not yet been
resolved in the context of string theory (although some interesting
partial results have been obtained \cite{lawr:sing}).

A related issue concerns the description of spacetime inside the event
horizon.  In the original calculations of the entropy of extreme and
near-extreme black holes in string theory
\cite{strom:D-ent,cal:D-ent,hor:neD-ent}, one considered a
configuration of D-branes and counted the number of states of given
energy at weak coupling. (These excitations are described by a
supersymmetric gauge theory.) One then extrapolated to strong coupling
and compared this number with the area of the event horizon of the
corresponding black hole. Since the area of the event horizon is much
less than the string scale at weak coupling, it wasn't clear what the
relation was between the quantum states of the gauge theory and the
region of spacetime behind the horizon. There was speculation that
perhaps the branes were located near the horizon, replacing the region
inside \cite{cal:D-ent}, or that the description of the
gauge theory interacting with the bulk supergravity modes was
appropriate only for outside observers. Another description would then
be needed for infalling observers.

Recent developments in string theory have shed light on both of these
issues.  It was proposed in \cite{juan:N1} that for large $N$ and
strong coupling ($g^2_{YM}N\gg 1$), supersymmetric $U(N)$ gauge
theories are equivalent to string theory in particular supergravity
backgrounds. (For earlier work along these lines, see
\cite{kleb:3braneabs,gub:3braneabs}.) The basic idea is to consider
$N$ D-branes in string theory in a low energy limit. From an effective
field theory analysis, one argues that the degrees of freedom on the
branes decouple from the bulk modes in this limit. Starting with the
supergravity solution describing the branes and applying the same
limit, one obtains a region of spacetime near the horizon. Since one
starts with the same theory and applies the same limiting procedure,
the resulting descriptions should be equivalent. This bold conjecture
has been clarified in \cite{gub:corr,witten:eucl} and investigated in
a number of recent papers
\cite{juan:N2,hor:N,kachru:orb,kallosh:conf,berkooz:1,rey:quarks,juan:wilsonl,lawr:cft,gub:fixed,castel:g/h,gomis:inter,claus:conf,fer:sing1,fer:sing2,fer:sing3,flato:sing,aharony:m,minwalla:m},
with promising results. In this paper, we will see that if we follow
the argument of \cite{juan:N1} for near-extreme branes, we find that
the gauge theory at finite energy density is equivalent to string
theory in an asymptotically anti-de Sitter black hole solution,
including the region behind the horizon.

Let us first review the argument in the extreme case. To be concrete,
we consider $N$ D-three branes in type IIB string theory. The
supergravity solution has constant dilaton and metric 
\begin{equation}\label{brane}
ds^2 = f^{-1/2} ( -dt^2 + dy_i dy^i) + f^{1/2}(dr^2 + r^2
d\Omega_5),
\end{equation}
with
\begin{equation}\label{fdef}
f(r) = 1 + {4\pi gN \alpha'^2\over r^4},
\end{equation}
where $g$ is the string coupling ($g\sim g_{YM}^2$), and $y_i, \
i=1,2,3$ denote the directions along the branes. The horizon is at
$r=0$, and these coordinates (with $r>0$) cover only the region
outside the horizon. The maximal analytic extension contains an
infinite number of asymptotically flat regions, but no singularities
\cite{gib:brane}. The low energy limit consists of setting $r = U
\alpha'$, and taking $\alpha' \rightarrow 0$ with $U$ and $g$ fixed. This
effectively restricts attention to the region near the horizon.  The
resulting metric is
\begin{equation}\label{nearhor}
ds^2 = {U^2\over \sqrt{4\pi gN}} (-dt^2 + dy_idy^i)
+ {\sqrt{4\pi gN}\over U^2} dU^2 + \sqrt{4\pi gN} d\Omega_5
\end{equation}
in string units. This metric describes the product of $S^5$ and five
dimensional anti-de Sitter space ($AdS_5$), both with radii $R^2 =
\sqrt{4\pi gN}$. The gauge theory associated with $N$ D-three branes
is four dimensional $U(N)$ ${\cal N} = 4$ super Yang-Mills. The
conjecture is that the strong coupling limit of this gauge theory is
equivalent to string theory on $AdS_5\times S^5$, with the above
radii.

In \cite{hor:N} it was argued that the gauge theory actually describes
string theory on both sides of the horizon $U=0$. This is because the
gauge theory has a conformal symmetry group $SO(2,4)$, which is
supposed to be identified with the isometry group of the near horizon
geometry. $SO(2,4)$ is indeed the symmetry group of $ AdS_5$, since
$AdS_5$ is the surface $-UV + \eta_{\mu\nu}X^\mu X^\nu = -R^2$ in a
flat spacetime with signature $(2,4)$, ${\bf R}^{(2,4)}$. However, the
region outside the horizon ($U>0$) is not invariant under this full
group, but only under a subgroup which leaves a null direction in
${\bf R}^{(2,4)}$ invariant.  So the gauge theory, which is invariant
under the full group, must describe string theory on both sides of the
horizon.  Furthermore, to properly realize the conformal symmetry, the
gauge theory should not be defined on ${\bf R}^4$ but rather on $S^3
\times {\bf R}$. Time evolution on $S^3 \times {\bf R}$ corresponds in
$AdS_5$ to evolution with respect to a global timelike symmetry which
does not leave the region $U>0$ invariant.

\sect{Near-extreme D-three branes}

\begin{figure}
\leavevmode
\centering
\epsffile{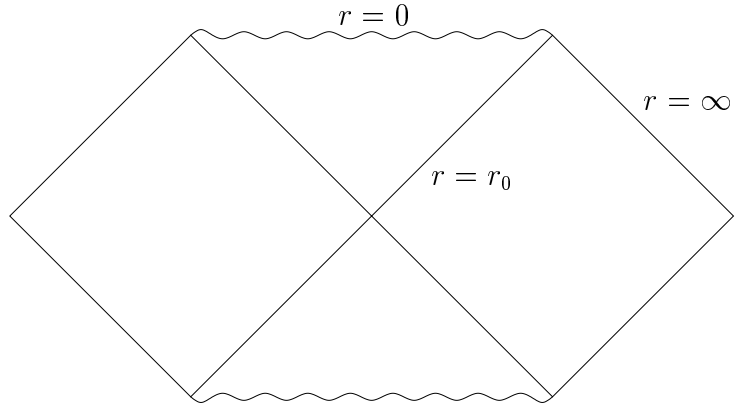}
\caption{Causal structure for the near-extreme 3-brane. Each point
represents $S^5 \times R^3$.
\label{fig1}}
\end{figure}

The argument of \cite{hor:N} is encouraging, but does not address the
issue of black hole singularities, since the extremal three brane is
regular everywhere. The horizons in the $AdS_5$ geometry are also not
true event horizons, since they depend on a particular choice of
Killing field.  For the near-extremal three brane, the metric is given
by
\begin{equation}\label{nexbrane}
 ds^2 = f^{-1/2} ( -h dt^2 + dy_i dy^i) + f^{1/2}(h^{-1} 
dr^2 + r^2 d\Omega_5),
\end{equation}
where $f$ is given by (\ref{fdef}) and 
\begin{equation} \label{hdef}
h(r) = 1-{{r_0}^4\over r^4}.
\end{equation}
The global structure of this solution is similar to the Schwarzschild
solution, and is shown in Fig. \ref{fig1}. This nonextremal solution
corresponds to excited D-branes whose dynamics are now described by
the gauge theory at a finite energy density.

\begin{figure}
\leavevmode
\centering
\epsffile{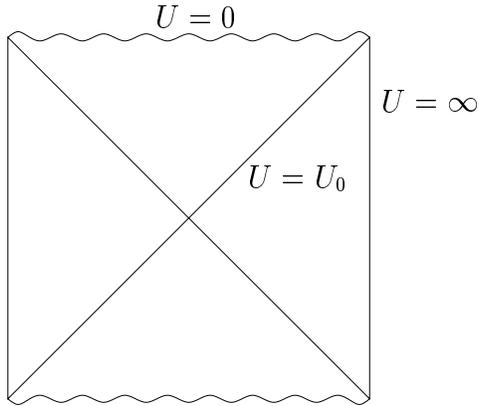}
\caption{Causal structure for the decoupled region.
\label{fig2}}
\end{figure}

One might have expected that the limit which decouples the gauge
theory from the bulk would restrict to the region near the horizon,
and both the singularity and the asymptotically flat region would be
removed. This is not the case. In fact one obtains the metric
\begin{equation} \label{bhmet}
ds^2 = {U^2\over \sqrt{4\pi gN}} 
\left[-\left(1-{U_0^4\over U^4}\right) dt^2 + dy_idy^i\right]
+ {\sqrt{4\pi gN}\over U^2}\left(1-{U_0^4\over U^4}\right)^{-1} dU^2 
+ \sqrt{4\pi gN} d\Omega_5.
\end{equation}
Note that the five sphere decouples throughout the spacetime
(\ref{bhmet}).  The Penrose diagram for the remaining five dimensional
spacetime is shown in Fig. \ref{fig2}. This is similar to the
Schwarzschild anti-de Sitter metric, but the symmetries are different;
there are translational invariances along the $y^i$ directions rather
than the spherical symmetry of Schwarzschild anti-de Sitter. The five
dimensional spacetime is asymptotically $AdS_5$, but it is not locally
$AdS_5$. While the Ricci tensor is constant $R_{\mu\nu} = -(4\pi
gN)^{-1/2} g_{\mu\nu}$, the Weyl tensor is nonzero, and given by
\begin{equation} \label{curv}
C_{ijij} \sim {U_0^4 \over U^4 \sqrt{4 \pi gN}},
\end{equation}
where the indices $i,j$ denote components with respect to an
orthonormal basis in the five-dimensional metric, and there is no sum
on $i,j$. The curvature clearly diverges as $U \rightarrow 0$.  The
five dimensional spacetime in (\ref{bhmet}) is a solution to
Einstein's equation with a negative cosmological constant, and is
analogous to four-dimensional metrics found earlier \cite{mann:top}.

Since the near extremal solution just corresponds to the gauge theory
at finite energy density ($\rho \sim U_0^4$), its clear that the gauge
theory can describe string theory in backgrounds other than anti-de
Sitter space. This should not be surprising, since the modes of the
string include fluctuations of the metric. Furthermore, since the
entire spacetime shown in Fig. 2 arises in the scaling limit, the
gauge theory must describe string theory on this entire space.

One might worry that the gauge theory only captures the physics
outside the horizon, since the natural time on the worldbrane (for
${\bf R}^4$ topology) only covers the external region. However the
same argument could have been made in the extremal case, and we have
already seen that the gauge theory naturally includes the region
beyond the horizon in this case. It is true that the solution
(\ref{bhmet}) has less symmetry than $AdS_5$, so there is no direct
analog of the symmetry argument of \cite{hor:N} to guarantee that the
region inside the horizon is included. However, the gauge theory is
the same as in the extreme case; one is simply considering different
states. Also, the difference between $S^3$ and ${\bf R}^3$ should not
matter for reasonable energy densities.

Note that if we considered a Euclidean version of the solution, as in
\cite{witten:eucl}, we would find that the Euclidean section arises
from analytically continuing the region outside the horizon, and we
would expect the Euclidean version of the gauge theory to correspond
to string theory on this Euclidean manifold. This does not contradict
our Lorentzian argument. String theory is consistent on the Euclidean
solution since it is a complete space. String theory cannot be
consistently defined on the region outside the event horizon of the
Lorentzian solution, since strings will fall in.

The supergravity approximation will break down once the curvature
becomes large, or if there are compact directions smaller than the
string scale (so winding modes become light). For large $gN$, the
curvature is small near the horizon and everywhere outside.  If one
periodically identifies the $y_i$, the radii of these circles shrink
as one decreases $U$.  At small enough $U$, the circles will be
smaller than the string scale, and it would be more appropriate to
study the T-dual metric. However, since one can compactify the $y_i$
with arbitrary periods $L_i$, for a given energy density, one can
arrange for the circles to stay larger than the string scale until the
local curvature becomes large. Then the supergravity approximation
breaks down only near the singularity in the five-dimensional
spacetime.  Since the gauge theory is perfectly nonsingular, it must
provide a regular description of the spacetime, including the singular
region.

The near extremal three brane will Hawking radiate.  Since the finite
temperature gauge theory is clearly in equilibrium, it cannot describe
a time-dependent process such as the evaporation of the three brane.
Thus, the finite temperature gauge theory must describe the brane in
thermal equilibrium with a gas of string modes in anti-de Sitter
space. The negative cosmological constant acts like a confining box,
so unlike the asymptotically flat case, the thermal gas will have
finite energy.

\sect{Discussion}

We have argued that the analysis of \cite{juan:N1} for near-extreme
three branes implies that the finite temperature large $N$ gauge
theory is equivalent to string theory on a background spacetime which
describes a black hole in anti-de Sitter space. In particular, the
gauge theory describes the region inside the black hole's horizon,
and, since it is perfectly regular, provides a resolution of the
singularity inside the black hole. Similar results apply for the
near-extremal versions of the other cases discussed in \cite{juan:N1}.
For the two brane and five brane in M theory, the discussion in the
previous section is carried over essentially unchanged. For the
near-extremal black string obtained from a D1+D5 system in six
dimensions (and for five-dimensional black strings), the black hole
solution is just the BTZ black hole \cite{ban:2+1}. The geometry is
locally $AdS_3$ in this case, and there is no curvature singularity.
The near-extremal six dimensional black string does have a curvature
singularity, and it also has two horizons: an event horizon and
inner horizon. The decoupling limit includes both horizons, but not
the singularity.

If we take the $y_i$ to be periodically identified in (2.3), the
spacetime has two asymptotic boundaries, corresponding to the two
sides of the Penrose diagram in Fig.  \ref{fig2}. It seems likely that
the interpretation of this is the same as in general relativity: the
second asymptotic region is probably unphysical, and does not arise
when the process of formation is included.  For example, suppose one
starts with $N$ D three branes spread out over a ball in the
noncompact directions. This should act like a $q=m$ charged dust. If
the branes are given some initial inward radial velocity, they will
collapse to form the nonextreme black hole, but the Penrose diagram
will now have only one asymptotically $AdS$ region.

Work is underway to find further support for this remarkable
connection between gauge theory and black hole singularities. One
issue that evidently deserves further study is understanding the
causal structure of the spacetime from the gauge theory point of
view. That is, we would like to ask why a test three brane, say, is
restricted to move within the light cones of the metric
(\ref{bhmet}). At weak coupling, the causal dynamics of branes follows
from the Dirac-Born-Infeld action. Here we have only the super
Yang-Mills action, but its possible that the DBI action arises as a
low energy effective action for probe branes. Studying infalling
branes (as in \cite{juan:nebhD}) seems a sensible next step to
obtaining a greater understanding of the connection between gauge
theory and geometry.

\bigskip
\bigskip
\centerline{\bf Acknowledgments}
\medskip

It is a pleasure to thank the participants of the duality program at
the Institute for Theoretical Physics, Santa Barbara for
discussions. This work was supported in part by NSF grant PHY95-07065.

\begingroup\raggedright\endgroup


\begin{thebibliography}{10}
\newcommand{\enquote}[1]{``#1''}

\bibitem{strom:coni}
A.~Strominger, \enquote{Massless black holes and conifolds in string theory,}
  Nucl. Phys. {\bf B451}, 96 (1995), hep-th/9504090.

\bibitem{lawr:sing}
A.~Lawrence and E.~Martinec, \enquote{String field theory in curved space-time
  and the resolution of spacelike singularities,} Class. Quant. Grav. {\bf 13},
  63 (1996), hep-th/9509149.

\bibitem{strom:D-ent}
A.~Strominger and C.~Vafa, \enquote{Microscopic origin of the
  {B}ekenstein-{H}awking entropy,} Phys. Lett. {\bf B379}, 99 (1996),
  hep-th/9601029.

\bibitem{cal:D-ent}
C.~G. Callan, Jr. and J.~M. Maldacena, \enquote{D-brane approach to black hole
  quantum mechanics,} Nucl. Phys. {\bf B472}, 591 (1996), hep-th/9602043.

\bibitem{hor:neD-ent}
G.~T. Horowitz and A.~Strominger, \enquote{Counting states of near extremal
  black holes,} Phys. Rev. Lett. {\bf 77}, 2368 (1996), hep-th/9602051.

\bibitem{juan:N1}
J.~Maldacena, \enquote{The large {N} limit of superconformal field theories and
  supergravity,} hep-th/9711200.

\bibitem{kleb:3braneabs}
I.~R. Klebanov, \enquote{World volume approach to absorption by nondilatonic
  branes,} Nucl. Phys. {\bf B496}, 231 (1997), hep-th/9702076.

\bibitem{gub:3braneabs}
S.~S. Gubser, I.~R. Klebanov, and A.~A. Tseytlin, \enquote{String theory and
  classical absorption by three-branes,} Nucl. Phys. {\bf B499}, 217 (1997),
  hep-th/9703040.

\bibitem{gub:corr}
S.~S. Gubser, I.~R. Klebanov, and A.~M. Polyakov, \enquote{Gauge theory
  correlators from noncritical string theory,} hep-th/9802109.

\bibitem{witten:eucl}
E.~Witten, \enquote{Anti-de {S}itter space and holography,} hep-th/9802150.

\bibitem{juan:N2}
N.~Itzhaki, J.~Maldacena, J.~Sonnenschein, and S.~Yankielowicz,
  \enquote{Supergravity and the large {N} limit of theories with sixteen
  supercharges,} hep-th/9802042.

\bibitem{hor:N}
G.~T. Horowitz and H.~Ooguri, \enquote{Spectrum of large {N} gauge theory from
  supergravity,} hep-th/9802116.

\bibitem{kachru:orb}
S.~Kachru and E.~Silverstein, \enquote{4-{D} conformal theories and strings on
  orbifolds,} hep-th/9802183.

\bibitem{kallosh:conf}
R.~Kallosh, J.~Kumar, and A.~Rajaraman, \enquote{Special conformal symmetry of
  world volume actions,} hep-th/9712073.

\bibitem{berkooz:1}
M.~Berkooz, \enquote{A supergravity dual of a (1,0) field theory in six-
  dimensions,} hep-th/9802195.

\bibitem{rey:quarks}
S.-J. Rey and J.~Yee, \enquote{Macroscopic strings as heavy quarks in large {N}
  gauge theory and anti-de {S}itter supergravity,} hep-th/9803001.

\bibitem{juan:wilsonl}
J.~Maldacena, \enquote{Wilson loops in large {N} field theories,}
  hep-th/9803002.

\bibitem{lawr:cft}
A.~Lawrence, N.~Nekrasov, and C.~Vafa, \enquote{On conformal field theories in
  four-dimensions,} hep-th/9803015.

\bibitem{gub:fixed}
S.~S. Gubser, A.~Hashimoto, I.~R. Klebanov, and M.~Krasnitz, \enquote{Scalar
  absorption and the breaking of the world volume conformal invariance,}
  hep-th/9803023.

\bibitem{castel:g/h}
L.~Castellani {\em et~al.\/}, \enquote{{$G/H$ M}-branes and {$AdS_{(p+2)}$}
  geometries,} hep-th/9803039.

\bibitem{gomis:inter}
J.~Gomis, D.~Mateos, J.~Simon, and P.~K. Townsend, \enquote{Brane intersection
  dynamics from branes in brane backgrounds,} hep-th/9803040.

\bibitem{claus:conf}
P.~Claus, R.~Kallosh, J.~Kumar, P.~Townsend, and A.~V. Proeyen,
  \enquote{Conformal theory of {M2, D3, M5 and D1-branes + D5-branes},}
  hep-th/9801206.

\bibitem{fer:sing1}
S.~Ferrara and C.~Fronsdal, \enquote{Conformal {M}axwell theory as a singleton
  field theory on {$AdS(5)$, IIB} three-branes and duality,} hep-th/9712239.

\bibitem{fer:sing2}
S.~Ferrara and C.~Fronsdal, \enquote{Gauge fields as composite boundary
  excitations,} hep-th/9802126.

\bibitem{fer:sing3}
S.~Ferrara, C.~Fronsdal, and A.~Zaffaroni, \enquote{On {$N=8$ supergravity on
  $AdS(5)$ and $N=4$} superconformal yang-mills theory,} hep-th/9802203.

\bibitem{flato:sing}
M.~Flato and C.~Fronsdal, \enquote{Interacting singletons,} hep-th/9803013.

\bibitem{aharony:m}
O.~Aharony, Y.~Oz, and Z.~Yin, \enquote{{M} theory on {$AdS_p \times S^{11-p}$}
  and superconformal field theories,} hep-th/9803051.

\bibitem{minwalla:m}
S.~Minwalla, \enquote{Particles on {$AdS_{4/7}$} and primary operators on
  {M}$_{2/5}$ brane worldvolumes,} hep-th/9803053.

\bibitem{gib:brane}
G.~W. Gibbons, G.~T. Horowitz, and P.~K. Townsend, \enquote{Higher dimensional
  resolution of dilatonic black hole singularities,} Class. Quant. Grav. {\bf
  12}, 297 (1995), hep-th/9410073.

\bibitem{mann:top}
R.~B. Mann, \enquote{Pair production of topological anti-de {S}itter black
  holes,} Class. Quant. Grav. {\bf 14}, L109 (1997), gr-qc/9607071.

\bibitem{ban:2+1}
M.~Banados, C.~Teitelboim, and J.~Zanelli, \enquote{The black hole in
  three-dimensional space-time,} Phys. Rev. Lett. {\bf 69}, 1849 (1992),
  hep-th/9204099.

\bibitem{juan:nebhD}
J.~Maldacena, \enquote{Probing near extremal black holes with {D}-branes,}
  Phys. Rev. D {\bf 57}, 3736 (1998), hep-th/9705053.

\end{thebibliography}
\end{document}